\documentstyle[twocolumn,prl,aps]{revtex}
\begin{document}
\draft
\input{psfig}
\twocolumn[\hsize\textwidth\columnwidth\hsize\csname @twocolumnfalse\endcsname
\title{Exo-hydrogenated Single Wall Carbon Nanotubes }
\author{T. Yildirim$^{(1)}$, O. G\"{u}lseren$^{(1,2)}$, and S. Ciraci$^{(3)}$}
\address{$^{(1)}$ NIST  Center for Neutron Research,
National Institute of Standards and Technology, Gaithersburg, MD 20899 }
\address{$^{(2)}$ Department of Materials Science  and Engineering, 
University of Pennsylvania, Philadelphia, PA 19104}
\address{$^{(3)}$ Physics Department, Bilkent University, Ankara, Turkey}
\date{\today}
\maketitle

\begin{abstract}

An extensive first-principles study of fully exo-hydrogenated
zigzag (n,0) and armchair (n,n) single wall carbon nanotubes
(C$_n$H$_n$), polyhedral molecules including cubane, dodecahedrane,
and C$_{60}$H$_{60}$ points to crucial differences in the electronic
and atomic structures relevant to hydrogen storage and device applications.
C$_n$H$_n$'s  are estimated to be stable up to the radius of a (8,8)
nanotube, with binding energies  proportional to 1/R. Attaching a single
hydrogen to any nanotube is always exothermic. Hydrogenation of zigzag
nanotubes is found to be more likely than armchair nanotubes with similar
radius.
Our findings may have important implications for selective functionalization
and finding a way of separating similar radius nanotubes from each other.

\end{abstract}


]

\bigskip

Carbon nanotubes\cite{iijima} exhibit very unusual structural and
electronic properties,
suggesting a wide variety of technological applications~\cite{mint,review},
including the storage of hydrogen where the large effective surface
area promises a large absorption
capacity~\cite{h2str1,h2str2,h2str3,h2str4,h2str5,h2str6,h2str7}.
Unfortunately, the studies to date report conflicting results.
While some labs\cite{h2str1,h2str2} report hydrogen storage
densities up to 10 wt\% other labs report\cite{h2str3,h2str4} only
0.4 wt\% on the same system. Theories based on physisorption
have failed to predict such high uptake\cite{mdh2tube}.
To the best of our knowledge, studies of hydrogen chemisorption
in nanotubes are very limited\cite{tada,yuchen} 
 and are clearly needed to have a better
understanding of hydrogen  and nanotube system.  

\bigskip
\centerline{\psfig{figure=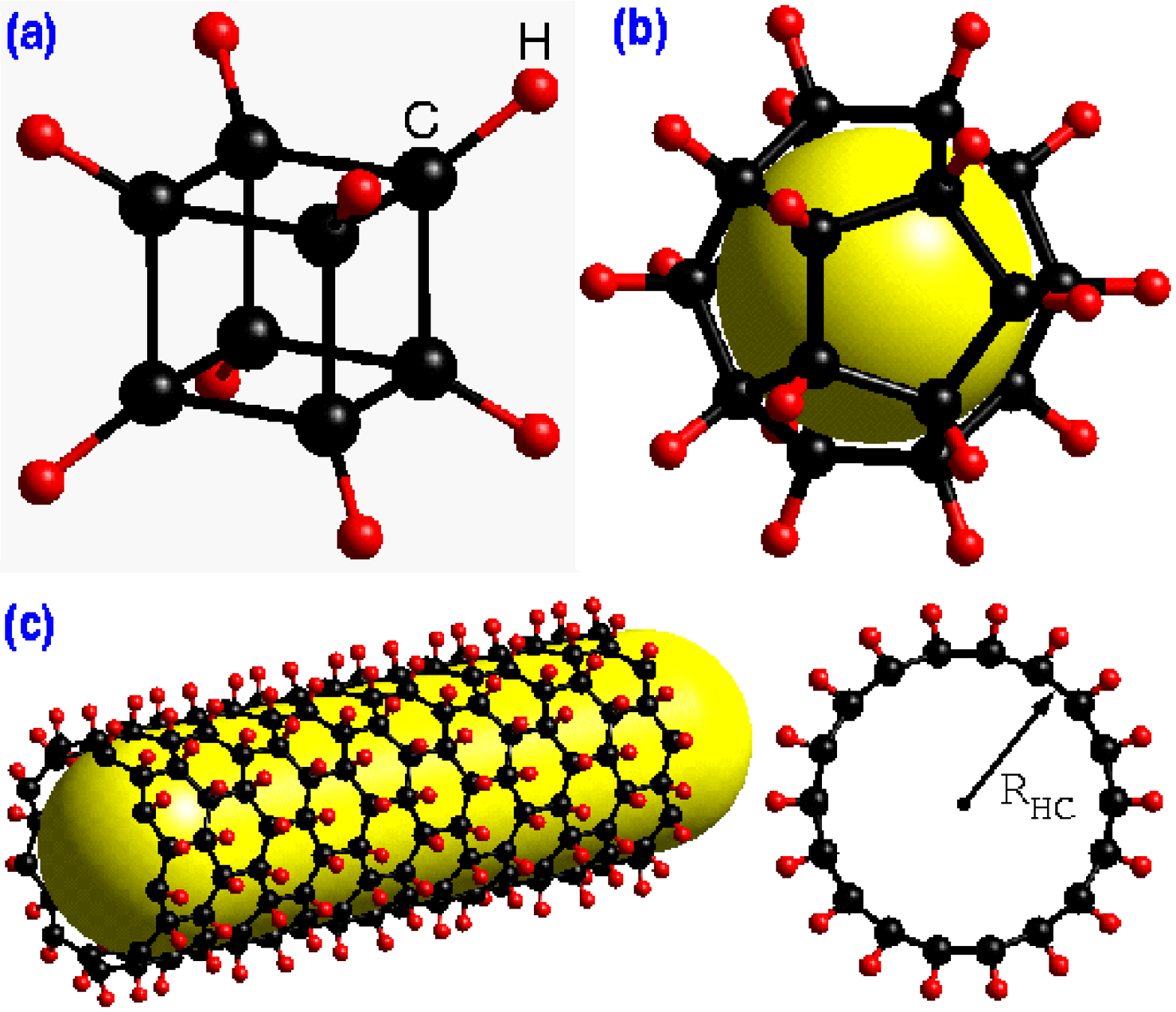,width=70mm}}

{\bf Fig.~1} {\small Three different polyhedra of carbon and 
hydrogen; (a) Cubane (O$_{h}$), (b) Dodecahedrane (I$_{h}$), and
(c) a side and top view of a single wall exo-hydrogenated carbon
nanotube. 
}

\bigskip

Hydrogen-carbon interactions have been studied extensively both 
theoretically and experimentally for many interesting polyhedral molecules,
such as cubane (C$_{8}$H$_{8}$)\cite{cubeaton,cubprl,cubprb}, 
dodecahedrane (C$_{20}$H$_{20}$)\cite{dodecahedrane}, and various
isomers of C$_{60}$H$_{n}$\cite{c60hx1,c60hx2} (see Fig.~1). 
Despite its very strained 90$^{\rm o}$ CCC-bond angle
cubane has been synthesized successfully\cite{cubeaton} (Fig.~1a).
Similarly, dodecahedrane and various isomers of C$_{60}$H$_{n}$
(up to n=32) have been also synthesized\cite{c60hx1}. 
These novel polyhedral molecules which represent the zero dimensional
case, exhibit many interesting properties. However, due to the one
dimensional nature and the curvature of carbon nanotubes, the
hydrogen-carbon interactions in these systems may be quite different
than those in polyhedral molecules. Therefore, it is important to know
if it is also possible to hydrogenate carbon nanotubes in a similar way
and if so what their structural and electronic properties would be.
This paper  addresses this important issue by performing extensive
first-principles calculations and shows that the chemisorption of
hydrogen is dependent on the radius and chirality of the nanotubes.
Theoretical predictions from first-principles studies played an important
role in guiding experimental studies in the
past~\cite{cubprb,interlink} and we expect that many findings
reported here may have important implications in this interesting system
as well. 

In order to obtain a reasonably complete understanding, we studied a
very large number of systems including zigzag (n,0) (n=7,8,9, 10 and 12)
and armchair (m,m) (m=4,5,6,8, and 10) nanotubes\cite{nmtubes}  
as well as cubane, 
dodecahedrane, C$_{60}$H$_{60}$ and finally hydrogenated graphene 
sheet (i.e. infinite limit of tube radius).
The first principles total energy and electronic structure
calculations were carried out using the pseudopotential 
plane wave method\cite{castep}. The results have been
obtained within the generalized gradient approximation (GGA)\cite{gga}.
This method has already been applied to many carbon 
systems, including fullerenes and cubane with remarkable
success\cite{cubprb,interlink}. 
We used  plane  waves with an energy cutoff of 500 eV. With this cutoff 
and using ultra soft pseudopotentials for carbon\cite{usp}, 
the total energy  converges within 0.5 meV/atom.
Interactions between molecules or nanotubes in periodic cells
are avoided by using large supercells. The supercell parameters are chosen
such that the closest H-H distance is 6 \AA. For molecular calculations
the Brillouin zone integration is carried out at the $\Gamma$-point.
For nanotubes, we used k-point spacing of $dk \approx 0.02 $ \AA$^{-1}$,
generating 5 and 10 special k-points along the tube-axis
for zigzag and armchair nanotubes, respectively\cite{kpts}. 
All carbon and hydrogen positions were relaxed without assuming any
symmetry. For nanotube calculations, the c-axis of the supercell
(corresponding to the tube axis) is also optimized. 

\bigskip
\centerline{\psfig{figure=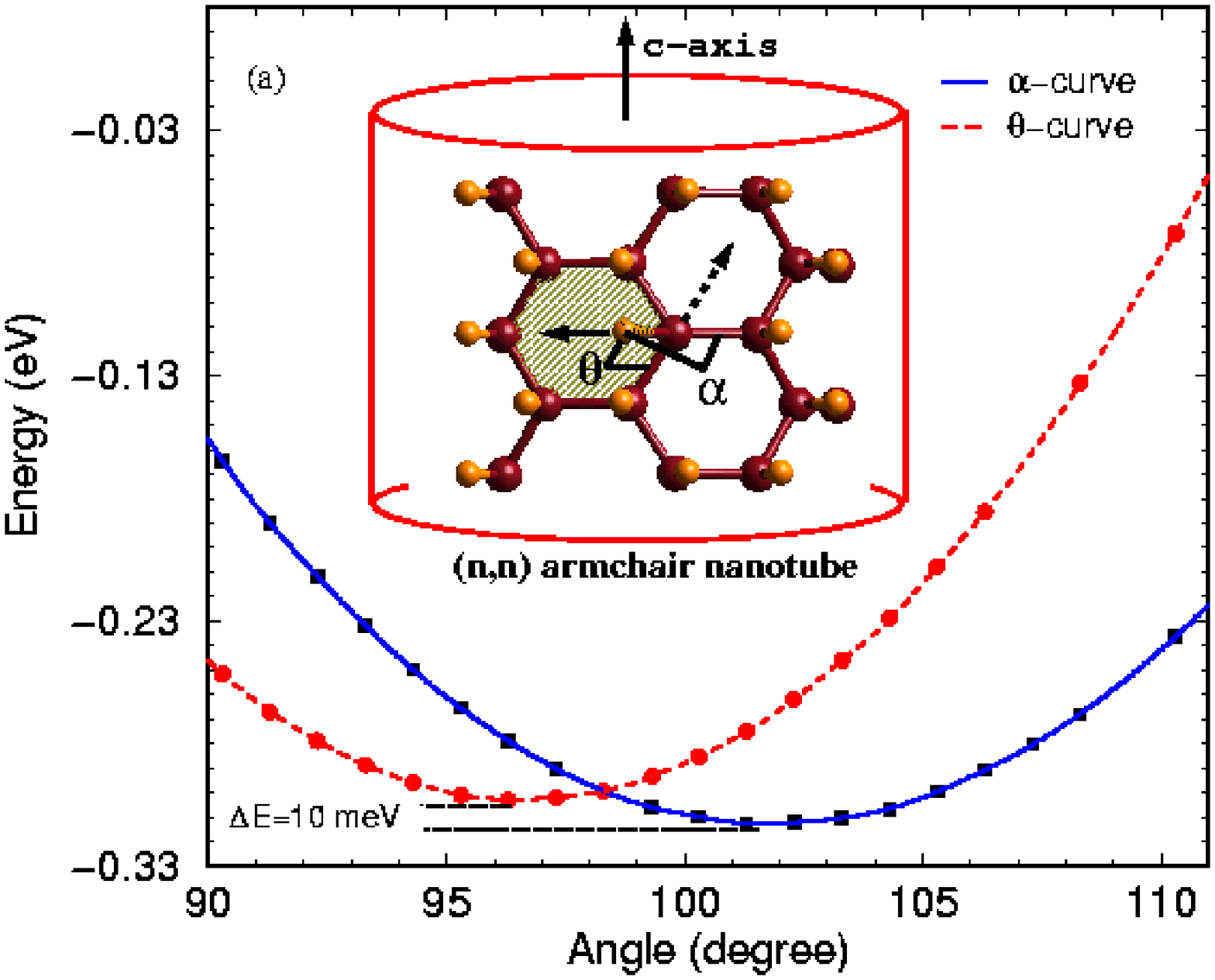,width=75mm}}

\centerline{\psfig{figure=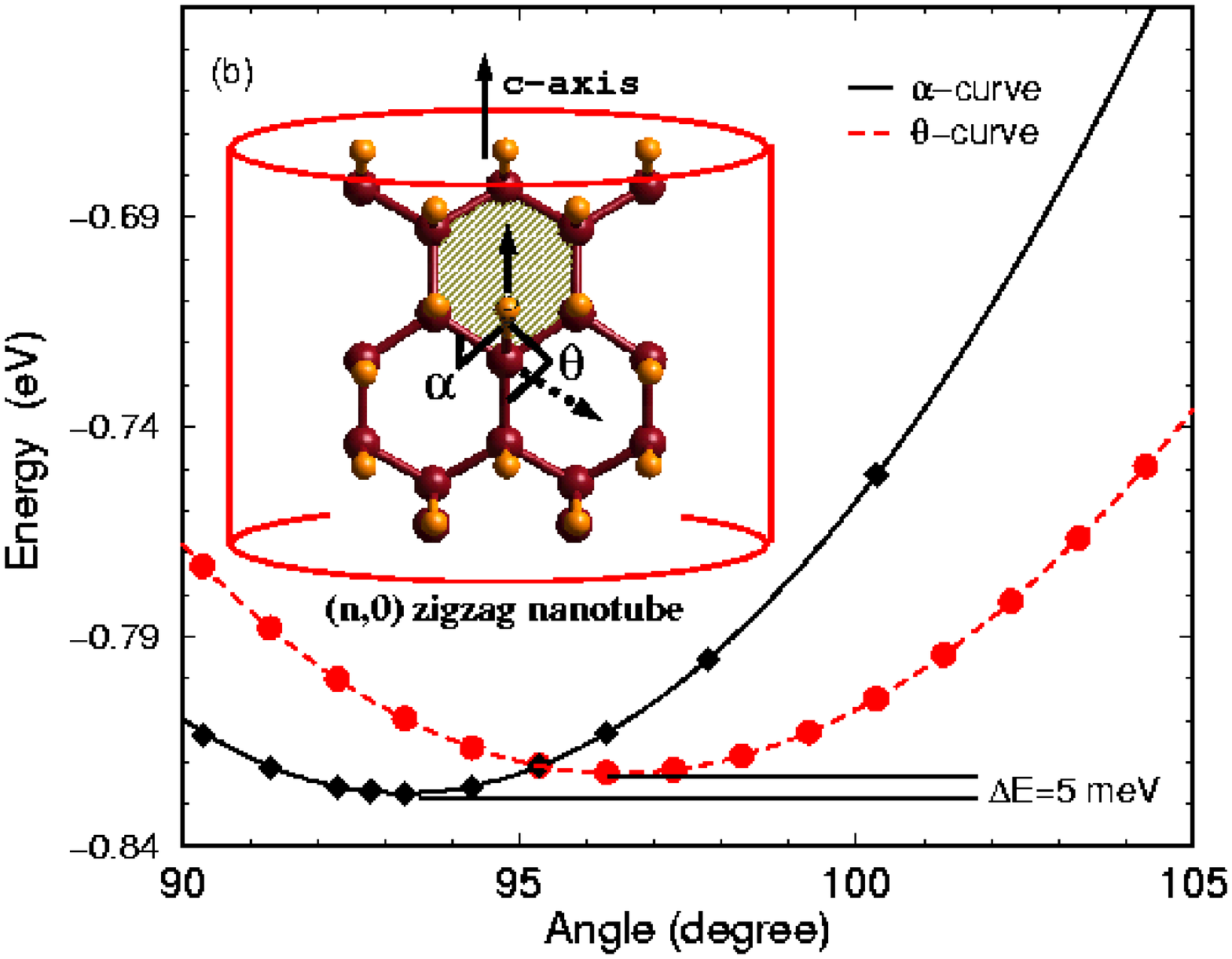,width=75mm}}

%
%
{\bf Fig.~2} {\small Energy curves as a CH-bond is rotated towards
the indicated arrows for armchair (top) and zigzag nanotubes (bottom)
starting from $\theta=\alpha$. The minimum energy is found when the
CH-bond is tilted  toward the shaded hexagons.
Zero of energy was taken to be arbitrary.}

\bigskip

In principle, there are an infinite number of isomers depending on the
locations of hydrogen atoms (i.e. endo if they are inside the tube and
exo if they are outside) as well as amount of hydrogen coverage.
Endo-hydrogenation, alternating endo-exo hydrogenation, and various half
coverage cases are being studied and the results will be published
elsewhere\cite{oguzcpl}. 
Here, we consider the case of full coverage where all carbon atoms
in a nanotube are hybridized with hydrogen atoms from outside
of the tube as shown in Fig.~1c. We refer to this isomer as a fully 
exo-hydrogenated carbon nanotube.  
 
First, the equilibrium orientations of the CH bonds were determined
starting with all the CH-bonds radially outward (Fig~1c).
Using this configuration, we studied 
a single CH-bond orientational dependence of the  potential energy surface.
Figure~2 shows the calculated energy curves as a single CH-bond is
rotated along two high symmetry directions for both zigzag and armchair
nanotubes. For armchair (n,n) nanotubes, the optimum orientation is
obtained when the CH-bond is tilted  about the tube axis (i.e c-axis).
Hence in the fully optimized structure, CH-bonds tilt in opposite directions
around the c-axis alternatively. For the zigzag nanotubes, the 
optimum orientation is obtained when the CH-bond is tilted towards the
c-axis. Therefore, the lowest energy configuration for zigzag tubes has
CH-bonds tilted towards plus and minus c-axis alternatively. Having located
the CH-bond orientations in this way, we next let all the carbon and
hydrogen atoms along with c-axis vary to obtain the final optimum structures. 

Table~1 summarizes the  parameters obtained for fully optimized structures.
Upon hybridization of carbons with hydrogens, the C-C bond length
($d_{CC}$) increases from  $ \approx  1.4 $ \AA $\;$ to $ \approx 1.55 $ \AA.
The latter is typical for sp$^{3}$ CC-bonds. The increase in 
$d_{CC}$ results in an increase in the tube radius (R$_{HC}$) 
by about  13 - 16 \%  for armchair nanotubes and by about 15 - 17\%
for zigzag nanotubes. Interestingly, these values are almost twice 
of those found for the polyhedral molecules.
Moreover, the value of $d_{CC}$ increases slightly
(by about 0.03 \AA) with increasing tube radius.
The CH-bond length ($d_{CH} \approx 1.09$ \AA) is also found to have
weak dependence on the tube radius. Using projection techniques we
estimated the total charge transfer from hydrogen to carbon to be around
0.26 electrons for nanotubes and 0.3 electrons for polyhedral molecules.

The most important difference between zigzag and armchair nanotubes is
found in the local CCH angles ($\alpha_{CCH}$). Even though one of these
angles is about the same for both types of nanotubes, the second angle in
zigzag nanotubes is always larger than those in armchair nanotubes.
This implies that the CCH-bond angles are more frustrated in armchair
nanotubes than in zigzag nanotubes and therefore deviate more from the 
ideal tetrahedral sp$^{3}$ bond angle of 109.5$^{\rm o}$.
This observation suggests that hydrogenated armchair nanotubes will have
higher energy and therefore they are less stable  than zigzag nanotubes.
Unlike CCH-bond angles, CCC-bond angles have weak radius dependence 
and are about the same for both types of nanotubes. 

\begin{minipage}[t]{17.5cm}

\begin{center}

\begin{tabular} {| c | c || c|  c|c|c|c| }
\hline \hline
Material & Formula  & R$_{HC}$ ( R$_{C}$) (\AA) &  d$_{CH}$ (\AA) & d$_{CC}$ (\AA)  & 
$\alpha_{CCH}$ (deg.) & $\alpha_{CCC} $ (deg.)   \\  \hline  \hline
(4,4) & C$_{16}$H$_{16}$   & 3.103 (2.734)     &  1.090  & 1.541, 1.567 & 96.70, 98.60 & 112.77, 120.69  \\ 
(5,5)& C$_{20}$H$_{20}$     & 3.885 (3.394)     &  1.087  & 1.549, 1.575 & 94.82, 97.15 & 113.18, 121.50  \\ 
(6,6)& C$_{24}$H$_{24}$     & 4.698 (4.061)     &  1.084  & 1.557, 1.594 & 93.35, 96.30 & 113.30, 122.00  \\ 
(8,8) & C$_{32}$H$_{32}$    & 6.228 (5.400)     &  1.079  & 1.567, 1.594 & 92.16, 94.85 & 114.62, 121.95  \\ 
(10,10)& C$_{40}$H$_{40}$   & 7.780 (6.755)     &  1.077  & 1.574, 1.600 & 91.40, 94.00 & 115.40, 121.76  \\ \hline
(7,0) & C$_{28}$H$_{28}$    & 3.180 (2.765)     &  1.092  & 1.549, 1.553 & 96.40, 102.25 & 113.95, 125.90  \\ 
(8,0)& C$_{32}$H$_{32}$     & 3.641 (3.146)     &  1.090  & 1.553, 1.557 & 95.22, 101.60 & 114.12, 127.00  \\ 
(9,0)  & C$_{16}$H$_{16}$   & 4.111 (3.557)     &  1.089  & 1.553, 1.566 & 94.32, 101.14 & 114.27, 127.58  \\ 
(10,0) & C$_{40}$H$_{40}$   & 4.571 (3.912)     &  1.087  & 1.556, 1.572 & 93.60, 100.54 & 114.48, 127.85  \\ 
(12,0) & C$_{48}$H$_{48}$   & 5.467 (4.695)     &  1.084  & 1.557, 1.576 & 92.66, 99.340 & 115.11, 127.67   \\ \hline
Cubane & C$_{8}$H$_{8}$  & 1.345 (1.267) &  1.087  & 1.553 & 125.26 & 90.0  \\ 
Dodecahedrane & C$_{20}$H$_{20}$  & 2.157 ( 2.0)  &  1.090  & 1.539 & 110.9 & 108.0  \\ 
Fullerene & C$_{60}$H$_{60}$  & 3.827 ( 3.510)  &  1.088  & 1.536 1.550 & 101.1, 101.9 & 108.1, 120.0  \\ 
\hline \hline
\end{tabular}

\end{center}

\noindent
{\bf Table 1 }{\small
Various  parameters  of the fully optimized structures of exo-hydrogenated
armchair and zigzag carbon nanotubes and other polyhedral molecules.
For graphene (i.e. $R_{HC} \rightarrow \infty$)
d$_{CH}$ and d$_{CC}$ are  1.066 \AA $\;$ and 1.622 \AA, respectively.
}
\end{minipage}

\bigskip

The stability and energetics of CH-bond formation are derived from the
average binding energy per atom for exo-hydrogenated nanotubes defined as 
\begin{equation}
E_{B} = ( E_{C_{n}H_{n}} - E_{C} - n E_{H} )/n .  
\end{equation}
Here  $E_{C_{n}H_{n}}, E_{C},$ and $E_{H}$ are the total energies
of the fully optimized exo-hydrogenated nanotube, nanotube alone, and
hydrogen atom, respectively. According to this definition, a stable
system will have a negative binding energy. Fig.~3a shows the radius
dependence of  $E_{B}$ for nanotubes and polyhedral molecules (see inset).
Two interesting observations are apparent. First, as shown by solid and
dotted lines, the binding energies can be very well described by a one
parameter fit;
\begin{equation}
E_{B} = E_{0} -C(n,m)/R_{HC}, 
\end{equation}
where $E_{0}$ is the limit $R_{HC} \rightarrow \infty $ (i.e. 
graphene) and calculated to be -1.727 eV. The fit results for 
$C(n,m)$ are given in Fig.~3a for zigzag and armchair nanotubes.
The inset to Fig.~3a shows that while E$_{B}$ for cubane falls on the
same curve as nanotubes, dodecahedrane and C$_{60}$H$_{60}$ have
lower energies than nanotubes due to the their more spherical shape.

The second interesting observation in Fig.~3a
is that the binding energies of zigzag nanotubes are always lower than
those in armchair nanotubes with similar radius by about 30 meV/atom.
As discussed above, this is a natural result of the fact that the
CCH-bond angles in zigzag nanotubes are closer to the optimum tetrahedral
sp$^{3}$ bonding than those in armchair nanotubes.
We expect this observation is also valid for hybridization of nanotubes
with other elements, such as Cl and F and this may have important
implications for separating similar radius nanotubes from
each other by selective chemical functionalization.

.
\vspace*{6.5cm}
.

\bigskip
\centerline{\psfig{figure=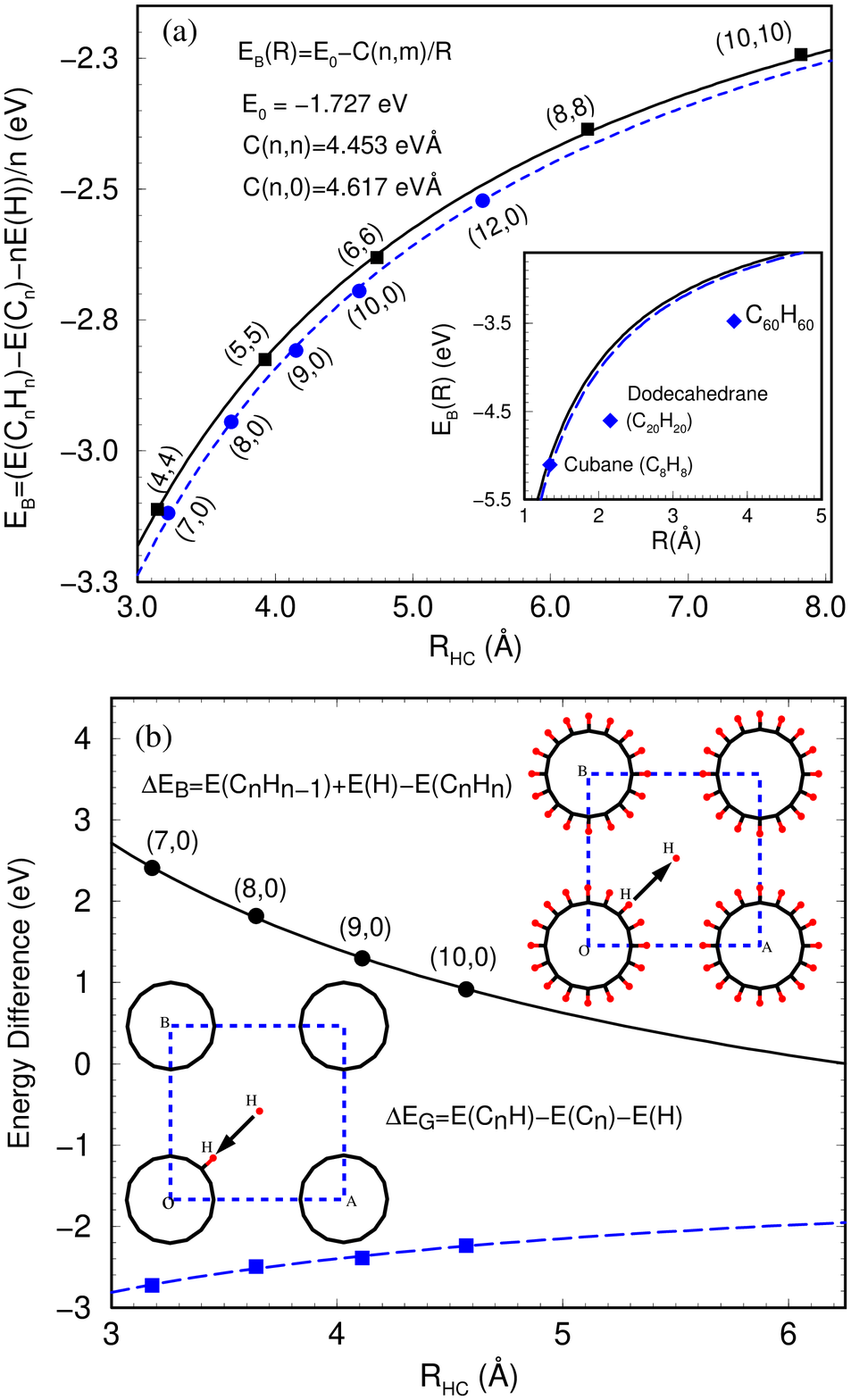,width=70mm}}

{\bf Fig.~3} {\small
(a) Binding energies $E_{B}$ of (n,n) (square) 
and (n,0) (circle) nanotubes as a function of R$_{HC}$. The solid and
dashed lines are one parameter fits to $ E_{B} = E_{0} -C(n,m)/R_{HC}$
as discussed in the text. Inset shows the binding energies of cubane,
dodecahedrane, and C$_{60}$H$_{60}$.
(b) Full circles indicate  the energy   ($\Delta E_{B}$) to break a single
CH-bond to form a  C$_{n}$H$_{n-1}$ zigzag nanotubes as depicted  in the
top inset. Full squares indicate energy gain $\Delta E_{G}$ by attaching
a single H atom to a nanotube to form a C$_{n}$H as depicted in the bottom
inset. The solid and dashed lines are two parameter fits as discussed in
the text, indicating 1/R$_{HC}$ behavior.}

\bigskip

Even though C$_{n}$H$_{n}$ nanotubes are found to be stable 
with respect to a pure carbon nanotube (C$_{n}$) and $n\times H$ atoms
for all values of the radius, it is of interest to see if they are also
stable against breaking a single CH-bond. We, therefore, calculated
energies of fully optimized hydrogenated nanotubes after breaking one
of the CH-bonds and putting the H atom at the center of supercell as
shown in the  top inset to Fig.~3b. 
Calculated values of the energy differences $\Delta E_{B}$,
for zigzag nanotubes were fitted to
$\Delta E_{B} = E_{0} + A/R_{HC}$ where
$E_{0}$ and $A$ are  2.506 eV and -15.671 eV\AA, respectively.
We note that for radius around $R_{HC} \approx 6.25 $ \AA, the
$\Delta E_{B}$  becomes negative, suggesting instability~\cite{barrier}.
Hence, (12,0) and (8,8) nanotubes are  at the limit for stable, fully
exo-hydrogenated nanotubes.
We are currently studying this
problem for half-coverage case as well. 

\bigskip
\centerline{\psfig{figure=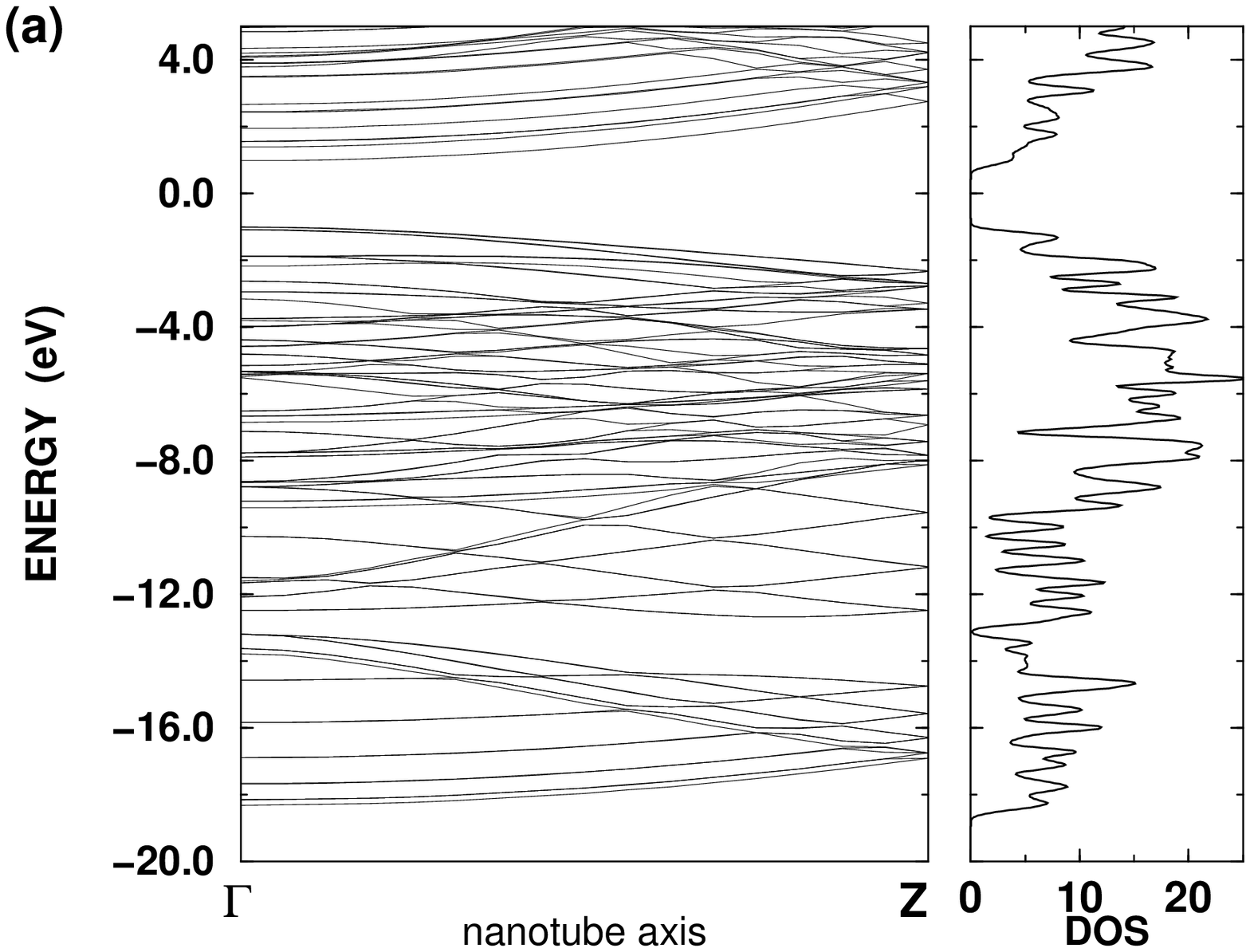,width=70mm}}

\centerline{\psfig{figure=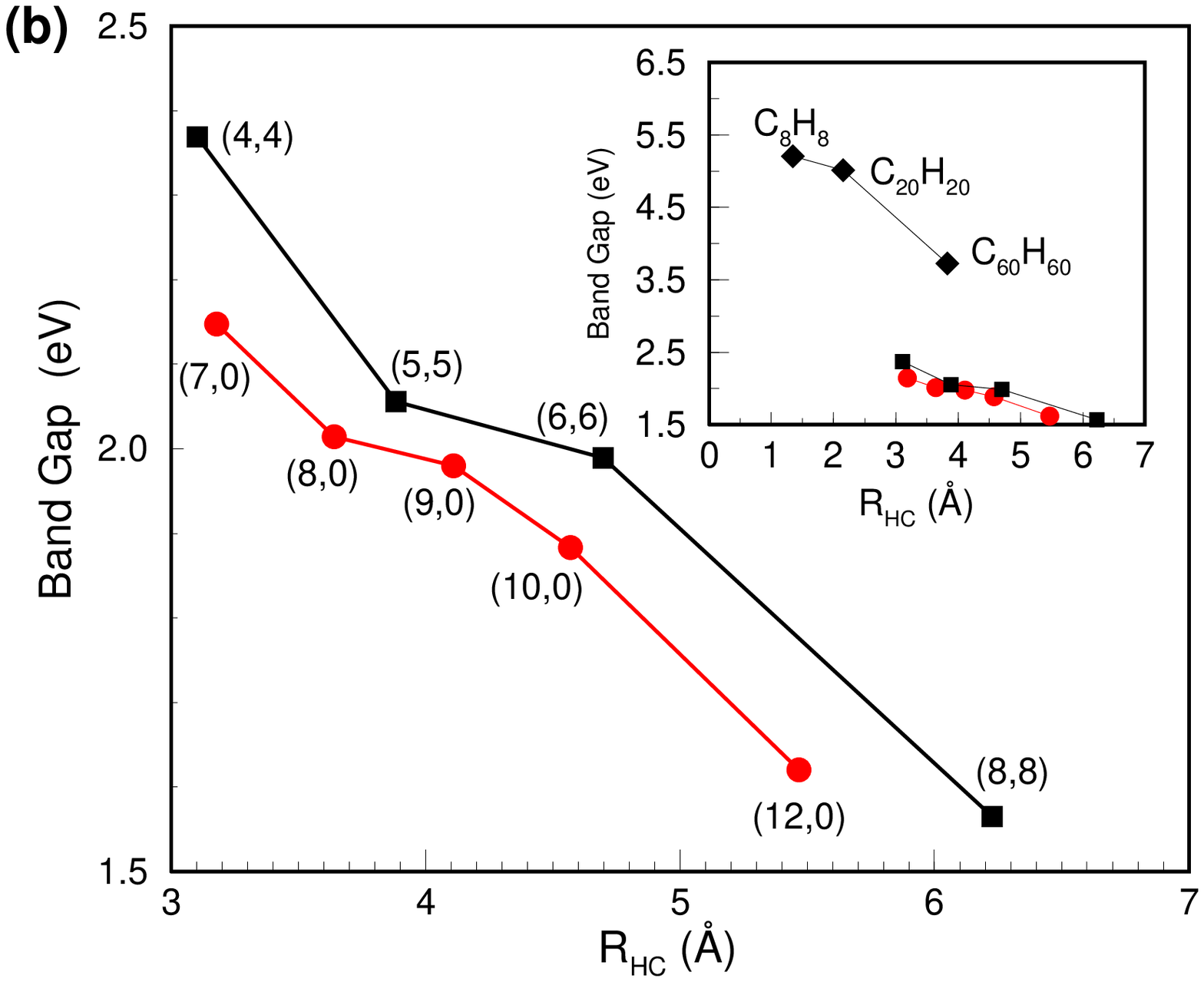,width=70mm}}

{\bf Fig.~4} {\small
(a) Electronic band structure  of a (9,0)
exo-hydrogenated carbon nanotube (left panel) and the corresponding
density of states (DOS) (right panel). (b) Band gap as a function
of tube radius R$_{HC}$. Inset shows the full scale plot to include 
the band gaps of the polyhedral molecules. }

\bigskip

The energy, $\Delta E_{G}$,  gained 
by attaching a single H atom to a carbon nanotube is calculated by 
performing structure optimization of a C$_{n}$H-nanotube 
as depicted  in the bottom inset to Fig~3b.
It is seen that $\Delta E_{G}$ can be
also well described by $\Delta E_{G} = E'_{0} + A'/R_{HC}$ (dashed line)
where $E'_{0}$ and $A'$ are  -1.161 eV and -4.952 1 eV\AA, respectively.
Unlike $\Delta E_{B}$, there is no change in the sign of $\Delta E_{G}$, 
suggesting that for any radius of carbon nanotube hybridization of
a single carbon atom is always stable. However the energy
gain from two such  processes is  around  5--6 eV which is 
slightly less  than the dissosiation  energy of H$_{2}$, 6.65 eV.
Hence C$_{n}$ nanotube plus H$_{2}$ system is stable
against forming a C$_{n}$H$_{2}$ hydrogenated nanotube. Therefore, 
in order to realize the CH-bonding discussed here, one first has to break
H$_{2}$ molecules into hydrogen atoms, probably by using  
a metal catalyst or electrochemical techniques.

Hydrogenation of nanotubes is also important in the modification
of the electronic structure for device applications.
Figure~4 shows the band structure and the corresponding density of
states (DOS) for a (9,0) exo-hydrogenated nanotube, which is typical to 
other nanotubes that we studied. Using projected DOS, we find that   
the bottom of the conduction bands are mainly derived from hydrogen  
while the top of the valence bands are mainly carbon-origin. 
In contrast to pure nanotubes which are metal or semiconductors
depending on their structure, the C$_{n}$H$_{n}$ nanotubes are found
to be direct band insulators with a gap of 1.5--2 eV at the $\Gamma$-point.
This value is about one-third of those for the molecular polyhedrals,
indicating less stability of hydrogenated nanotubes than molecules
(Fig.~4b). The band gaps decrease with increasing tube radius but unlike
binding energies there is no apparent 1/R$_{HC}$ type behavior. 
Interestingly, the band gaps of armchair
nanotubes are higher in energy by about 0.2 eV than those in 
zigzag nanotubes. This  is surprising because the band gap is usually
higher for more stable saturated hydrocarbons.

The observed band gap opening via hydrogenation of nanotubes can be used for 
band gap engineering for device applications such as metal-insulator
heterojunctions. 
For example, various quantum structures can easily be realized on
an individual carbon nanotube, and the properties of these structures 
can be controlled by partial hydrogenation of carbon nanotubes. 
If the different  regions of a SWNT are covered with
hydrogen atoms, the band gap and hence the electronic structure
will vary along the axis of the tube. This way various quantum 
structures of the desired size and  electronic character 
can be formed. In this respect, present scheme is quite similar to our 
previous constructions of nanotube heterostructures or quantum dots, 
where periodic applied transverse compressive
stress is used for band gap opening\cite{kilic}.

In summary, we have presented first-principles calculations of the
structural and electronic properties of various nanotubes which are
fully protonated by sp$^{3}$ hybridization of carbons.
We find that C$_{n}$H$_{n}$ nanotubes are stable for tube radius
R$_{HC}$ smaller than 6.25  \AA, roughly corresponding to a (8,8) nanotube.
Hybridization of a single carbon atom is found to be always exothermic
regardless of tube radius. Weak but stable CH-bonding in nanotubes may be
an important consideration for possible hydrogen storage applications. We also
found that hybridization of zigzag nanotubes is more likely than armchair
nanotubes with the same radius, suggesting a possible selective chemical
functionalization of nanotubes.
The fact that other carbon clusters such as cubane,
dodecahedrane, and C$_{60}$H$_{32}$ have been synthesized successfully,
suggest that it may  possible in the near future
to hydrogenate carbon nanotubes, yielding new structures
with novel properties.

{\bf ACKNOWLEDGEMENTS}.  
This work  is partially supported by the National Science
Foundation under Grant No. INT97-31014 and
T\"{U}B\.{I}TAK under Grant No. TBAG-1668(197 T 116).


\begin{references}
\bibitem{barrier}
There is an energy barrier for breaking a single CH-bond process
from the fully hydrogenated carbon nanotube, 
$\Delta E_{B}$ is the energy difference between the initial and
final states.
\end{references}
\end{document}